\documentclass[a4paper]{jpconf} 

\usepackage{graphicx}
\usepackage{iopams}
\usepackage{amsmath}

\newcommand{\exclude}[1]{}

\newcommand{\be}{\begin{equation}}
\newcommand{\ee}{\end{equation}}
\newcommand{\bea}{\begin{eqnarray}}
\newcommand{\eea}{\end{eqnarray}}

\def\+{\dagger}

\def\<{\langle}
\def\>{\rangle}

\def\l{\left}
\def\r{\right}

    \newcommand{\Mpc}{{\rm Mpc}}

    \newcommand{\GeV}{{\rm GeV}}
    \newcommand{\Gauss}{{\rm Gauss}}

\newcommand{\dd}{d}
\newcommand{\tconf}{t}
\newcommand{\V}[1]{{\boldsymbol{#1}}}

\newcommand{\Vk}{\V{k}}

\newcommand{\Vq}{\V{q}}

\newcommand{\dln}{\dd\ln}
\newcommand{\dlna}{\dln{}a}

\newcommand{\dlnk}{\dln{}k}

   \renewcommand{\H}{\mathcal{H}}
   \newcommand{\Fdual}{\widetilde{F}}
   \newcommand{\A}{\mathcal{A}}
   \newcommand{\CS}{\mathcal{S}}
   
   \newcommand{\fN}{{f_N}}
   
\begin{document}

\title{Generation of helical magnetic fields from inflation}

\author{Rajeev Kumar Jain, Ruth Durrer and Lukas Hollenstein}

\address{D\'epartement de Physique Th\'eorique and Center for Astroparticle Physics,\\
Universit\'e de Gen\`eve, 24 Quai Ernest Ansermet, CH--1211 Gen\`eve 4, Switzerland}

\ead{Rajeev.Jain@unige.ch}

\date{\today}

\begin{abstract}
The generation of helical magnetic fields during single field inflation due to an axial coupling of the electromagnetic field to the inflaton is discussed. We find that such a coupling always leads to a blue spectrum of magnetic fields during slow roll inflation. Though the helical magnetic fields further evolve during the inverse cascade in the radiation era after inflation, we conclude that the magnetic fields generated by such an axial coupling can not lead to observed field strength on cosmologically relevant scales. 
\end{abstract}

\section{Introduction}\label{intro}

There exists persuasive observational evidence for large scale magnetic fields with typical strength of $\mu$Gauss in various cosmic structures from stars to galaxies and galaxy clusters~\cite{Kronberg:1993vk, Pentericci:2000mp, Battaglia:2008ex}. Recently, using Fermi and HESS datasets, an interesting lower bound of $B \geq 10^{-16}$ Gauss has been obtained on the strength of magnetic fields in the intergalactic medium~\cite{Neronov:1900zz}.

All such observations naturally lead to the question of the origin of these magnetic fields. Various mechanisms of magnetic field generation can be broadly classified into two categories: early universe or primordial and late universe or astrophysical. Among the former ones, magnetic fields produced during inflation are most interesting as an arbitrary spectrum for the magnetic fields can be produced and therefore can lead to sufficient amplitude on large scales to provide seeds for the observed fields in galaxies and clusters~\cite{Martin:2007ue,Subramanian:2009fu}.

In what follows, I shall briefly discuss the magnetic field generation during inflation due to an axial coupling and show that, contrary to a non-helical coupling, a helical coupling always leads to a magnetic spectral index $n=1$ during slow roll~\cite{Durrer:2010mq}.
We further show that the backreaction turns out to be negligible as the magnetic energy density at the end of inflation is too small.
Even though the inverse cascade in the radiation era after inflation transfers power from small to large scales, the resulting field strength on cosmologically interesting scales is still insufficient to provide seeds for the observed magnetic fields in cosmic structures~\cite{Durrer:2010mq}.

\section{Axial coupling and the Fourier mode equation} \label{sec:basics}

Since the standard electromagnetic (EM) action is conformally invariant, the EM field fluctuations are not amplified in the conformally flat expanding background of inflation. In order to generate magnetic fields, one needs to break the conformal invariance of the EM field, e.g.~by coupling the EM field to a scalar or a pseudo-scalar field or to a curvature invariant. Here, we investigate the first possibility and study a helical coupling given by 
\be
  \mathcal{L}_{I}(\phi,A_\mu) = \frac{1}{4} f(\phi) F_{\alpha\beta}\Fdual^{\alpha\beta} \,.
\ee
It describes a coupling of the scalar field $\phi$ or the inflaton to the parity-violating term, $F\Fdual$, where $\Fdual$ is the dual of the EM field tensor and is defined as $ \Fdual^{\mu\nu} \equiv \frac{1}{2}\eta^{\mu\nu\alpha\beta}F_{\alpha\beta}$ where $F_{\mu\nu}\equiv\partial_{\mu}A_{\nu}-\partial_{\nu}A_{\mu}$ and $\eta^{\mu\nu\alpha\beta}$ is the totally anti-symmetric tensor with $\eta^{0123} \equiv (-g)^{-1/2}$. 
Adopting the Coulomb gauge where $A^\mu=(0,A^i)$ with $\partial_iA^i=0$, we find that the inhomogeneous Maxwell equation with the axial current becomes
\be
\ddot A_i -\nabla^2 A_i = -f'(\phi)\dot\phi\, \epsilon_{ijk}\partial_j A_k \, ,
\ee
where $f'(\phi)\equiv df/d\phi$ and the overdots denote derivative with respect to the conformal time $\tconf$. 
Note that, for a constant axial coupling, $f'(\phi)=0$, the sourced Maxwell equation reduces to the standard free wave equation and no EM fluctuations are amplified during inflation.
After quantization of the vector potential, we find that the Fourier modes $\A_h(\tconf,k)$ corresponding to the helicity states $h=\pm$ satisfy the wave equation~\cite{Durrer:2010mq}
\be
\ddot\A_h +\left[k^2+hkf'(\phi)\dot\phi\right]\A_h = 0 \,.
\label{eq:modeeq}
\ee

The power spectrum of a primordial stochastic magnetic field, statistically homogeneous and isotropic, is given by two scalar functions
and can be written as
\be
 \langle \widetilde{B}_i(\tconf,\Vk) \widetilde{B}^*_j(\tconf,\Vq) \rangle
  = \frac{(2\pi)^3}{2} \delta(\Vk-\Vq)
  \Big\{(\delta_{ij} -\hat{k}_i\hat{k}_j) P_S(\tconf,k)
  - i \epsilon_{ijn} \hat{k}_n P_A(\tconf,k) \Big\},
\ee
where $\widetilde{B} \equiv a^2 B$ is the rescaled magnetic field by the expansion factor $a$ and $P_{S/A}$ are the symmetric/anti-symmetric parts of the power spectrum, respectively. With respect to the helicity basis, the spectra can directly be written as
\be\label{e:PSA}
  P_{S/A}(\tconf,k) = k^2\left( |\A_+(\tconf,k)|^2 \pm |\A_-(\tconf,k)|^2 \right) \,,
\ee
and the rescaled magnetic energy density per logarithmic wave number is given by 
\be
  \frac{\dd \widetilde{\rho}_B}{\dlnk}(\tconf,k) = \frac{k^3}{(2\pi)^2}
  P_S(\tconf,k)\,.
\ee

\section{The magnetic power spectra in slow roll inflation} \label{sec:slowroll}

In order to treat the theory perturbatively, the strength of the interaction between the scalar field and the EM field should be small at all times which then leads to $|\fN|<1$ where $\fN$ is the dimensionless part of the coupling term defined as $ \fN \equiv \dd f/\dlna = f'(\phi)\dot\phi/\H $. In the slow roll regime where $\fN$ is approximately constant, the mode equation~(\ref{eq:modeeq}) can be solved analytically in terms of Coulomb wave functions. By requiring the solution to approach the free wave solution as initial condition, the full normalized solution can be written as
\be
  \A_h(k,\tconf) = \frac{1}{\sqrt{2k}}\Big[ G_0(-h\fN/2,-k\tconf) + iF_0(-h\fN/2,-k\tconf) \Big].
\ee
Using the asymptotic behaviour of this solution at late times, we find that the power spectra are given by~\cite{Durrer:2010mq}
\be
P_S(k) = k\, \frac{\sinh(\pi\fN)}{\pi\fN}
\quad {\rm and} \quad
P_A(k) = k\, \frac{\cosh(\pi\fN)-1}{\pi\fN}\,.
\ee
Notice that both the spectra are proportional to $k$ and only their amplitude changes with the coupling strength $\fN$ which tend to $(2\pi\fN)^{-1}\exp(\pi\fN)$ for large values of $\fN$. Using the symmetric power spectrum, the magnetic energy density per logarithmic wave number at the end of inflation can directly be computed to be
\bea \label{e:rhoM}
\frac{\dd \rho_B}{\dlnk}(\tconf_{\rm end},k) = \frac{k^4}{a_{\rm end}^4}\,\frac{\sinh(\pi\fN)}{4\pi^3\fN}
\equiv \frac{k^4}{a_{\rm end}^4}\,\CS^2(\fN)\,,
\eea
which then leads to 
\be
  \rho_B(\tconf_{\rm end}) \simeq \frac{1}{4}H_{\rm end}^4\,\CS^2(\fN) \,,
\ee
where $H_{\rm end}$ is the Hubble parameter at the end of inflation. Now, using the Friedmann equation, we can write
\be
  \Omega_B(\tconf_{\rm end}) \equiv \frac{\rho_B}{\rho_{\rm tot}}
  \simeq \frac{\CS^2(\fN)}{12} \left(\frac{H_{\rm end}}{m_P}\right)^2\,.
\ee
The generic condition for backreaction to be negligible is $\Omega_B < 1$ which then leads to
\be\label{e:back}
\CS(\fN) \lesssim \frac{m_P}{H_{\rm end}}\simeq \l(\frac{m_P}{T_*}\r)^2
\simeq 1.5 \times 10^{10} \left(\frac{10^{14}\,{\rm GeV}}{T_*}\right)^2\,,
\ee
where $T_*$ is the reheating temperature, given by $T_*^4 \simeq m_P^2H_{\rm end}^2$. In terms of $\fN$, the backreaction bound becomes
\be
 \fN \lesssim 17 \left(\frac{10^{14}\,{\rm GeV}}{T_\ast}\right).
\ee
Therefore, no backreaction on the background evolution is expected if inflation ends well below the Planck scale~\cite{Durrer:2010mq}.
Typically, the effective coupling term can be large in a situation wherein the EM field is coupled to a large number of pseudo-scalar fields~\cite{Anber:2006xt}. 

\section{Final strength of the magnetic fields}\label{sec:interpretation}

\begin{figure}[t]
\vskip -15pt
\begin{center}
\includegraphics[width=0.485\textwidth]{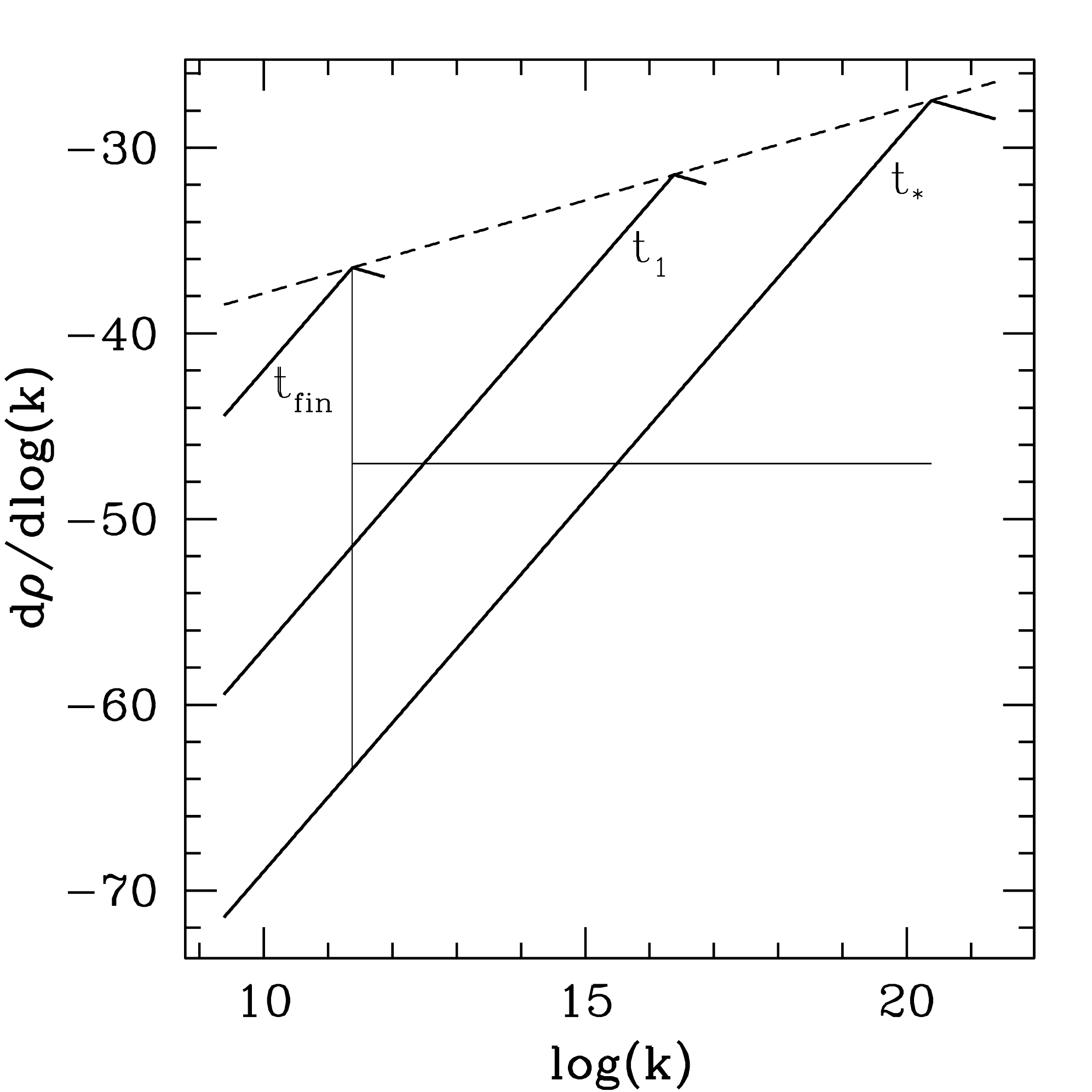}
\caption{The process of inverse cascade is depicted. The horizontal line indicates the total increase in the correlation scale during the inverse cascade while the vertical line represents the total amplification factor. On large scales, the shape of the spectrum remains unchanged while the amplitude increases because the correlation scale grows and the total helicity is conserved.}
\label{fig:invcascade}
\end{center}
\end{figure}

After inflation, due to helicity conservation, the magnetic field evolves by inverse cascade which can move power from small scales to large scales without affecting the spectral shape of the spectrum on large scales~\cite{Campanelli:2007tc}. On the other hand, on small scales, the magnetic fields are exponentially damped due to viscosity of the plasma. A sketch of the process of inverse cascade is shown in Fig.~\ref{fig:invcascade}.
Taking into account the maximal amplification of the magnetic field due to inverse cascade for our case, we find that the final strength of the magnetic field is given by
\be
\widetilde B(k)\ \simeq\ 3\times 10^{-19} \,\Gauss\; \CS(\fN)
 \left(\frac{k}{10^{12}/\Mpc}\right)^2 \left(\frac{T_*}{10^{14}\,{\rm GeV}}\right)^{9/11}.
\ee
At a typical scale of $\lambda=0.1\,\Mpc$ or $k=10/\Mpc$, the field strength becomes
\be 
 \widetilde B(k=10/\Mpc)\ \simeq\ 3\times 10^{-41} \,\Gauss\ \CS(\fN)
  \left(\frac{T_*}{10^{14}\,\GeV}\right)^{9/11}.
\ee
which is by far insufficient for subsequent dynamo amplification which requires seed fields of the order of at least
$10^{-20}\,\Gauss$~\cite{Brandenburg:2004jv}. Furthermore, to estimate how large this field can become for our best case, we require 
$\CS(\fN)<(m_P/T_*)^2$ and therefore
\be
  \widetilde B(k=10/{\rm Mpc})\ \le\ 10^{-32}\,\Gauss
    \left(\frac{10^{14}\,\GeV}{T_*}\right)^{13/11}.
\ee
In order to obtain the minimal necessary field for dynamo amplification of about $10^{-20}\,\Gauss$, we would require $T_* \le 10^4\,\GeV$ which corresponds to a rather low inflation scale, but not completely excluded~\cite{Durrer:2010mq}.

\section{Conclusions}\label{TheEnd}

We have discussed the generation of helical magnetic fields during inflation induced by an axial coupling term. During slow roll inflation, we have shown that the magnetic field power spectrum is always blue with spectral index $n=1$. Since the helical coupling is active only around horizon crossing and is nearly constant during slow roll, all modes are coherently amplified which leaves the shape of the  spectrum invariant with respect to the initial one. Although after inflation and reheating, an inverse cascade can transfer power from small to large scales, it turns out that for typical reheating temperatures, the magnetic fields are of ${\cal O}(10^{-40}\,\Gauss)$ which are largely insufficient for dynamo amplification~\cite{Brandenburg:2004jv}. 
Recently, we have shown that even the resonant amplification of magnetic fields due to an oscillating coupling fails to provide sufficient amplitude for the observed field strength due to extremely large backreaction~\cite{Byrnes:2011aa}.

\ack
The authors acknowledge financial support from the Swiss National Science Foundation.

\section*{References}
\bibliography{Inf-mf-refs}
\bibliographystyle{iopart-num}

\end{document}